\documentclass[aps,prl,10pt,superscriptaddress,nofootinbib,twocolumn]{revtex4-2}

\usepackage{amsmath,amssymb,times,hyperref,graphicx}
\usepackage[T1]{fontenc}
\hypersetup{colorlinks=true, linkcolor=blue, citecolor=blue, filecolor=blue, urlcolor=blue}

\newcommand{\ie}{\textit{i.e.}, }
\newcommand{\eg}{\textit{e.g.}, }
\newcommand{\cf}{\textit{cf.} }

\renewcommand{\d}{\mathrm{d}}
\newcommand{\f}{\mathrm{f}}
\renewcommand{\i}{\mathrm{i}}
\newcommand{\p}{\mathrm{p}}

\newcommand{\kt}{k_\mathrm{B}\mathbb{T}}
\newcommand{\grad}{\boldsymbol{\nabla}}
\newcommand{\sgn}{\,\mathrm{sgn}\,}

\newcommand{\vecF}{\mathbf{F}}
\newcommand{\vecf}{\mathbf{f}}
\newcommand{\vecI}{\mathbf{I}}
\newcommand{\vecR}{\mathbf{R}}
\newcommand{\vecr}{\mathbf{r}}
\newcommand{\vecV}{\mathbf{V}}
\newcommand{\vecv}{\mathbf{v}}

\newcommand{\bTa}{\mathbf{H}}
\newcommand{\bta}{\boldsymbol{\eta}}
\newcommand{\bXi}{\boldsymbol{\Xi}}
\newcommand{\bxi}{\boldsymbol{\xi}}

\begin{document} 

\title{Hydrodynamic memory in overdamped colloidal dynamics} 

\author{Benjamin Sorkin}
\email{bs4171@princeton.edu}
\affiliation{Princeton Center for Theoretical Science, Princeton University, Princeton, NJ 08544, USA}

\author{G\"unther Turk}
\affiliation{Princeton Materials Institute, Princeton University, Princeton, NJ 08544, USA}

\author{Howard A. Stone}
\email{hastone@princeton.edu}
\affiliation{Department of Mechanical and Aerospace Engineering, Princeton University, Princeton, NJ 08544, USA}

\begin{abstract}
    Micron-sized colloid particles diffusing through a fluid experience hydrodynamic inertial and memory effects, the latter causing velocity autocorrelation to decay as a power law. At the same time, many theoretical descriptions treat colloidal diffusion in a fluid using overdamped Langevin dynamics for its convenience, eliminating velocity, omitting inertia, but also ignoring the power-law memory. In this Letter, we show that hydrodynamic memory survives in the overdamped (colloid-inertialess) limit. By identifying the dimensionless parameter controlling the crossover between early- and late-time dynamics, we derive in closed form the overdamped Langevin equation in the presence of hydrodynamic memory. This provides a theoretical framework for realistically describing colloidal dynamics in a fluid, and establishes a rigorous basis for the positional memory observed in high-resolution experiments. Our theory predicts that hydrodynamic memory becomes increasingly pronounced for smaller particles and under stronger external forcing, and offers experimental probes for the crossover from conventional exponential relaxation to memory-dominated power-law dynamics.
\end{abstract}

\maketitle

\begin{figure}
    \centering
    \includegraphics[width=0.99\linewidth]{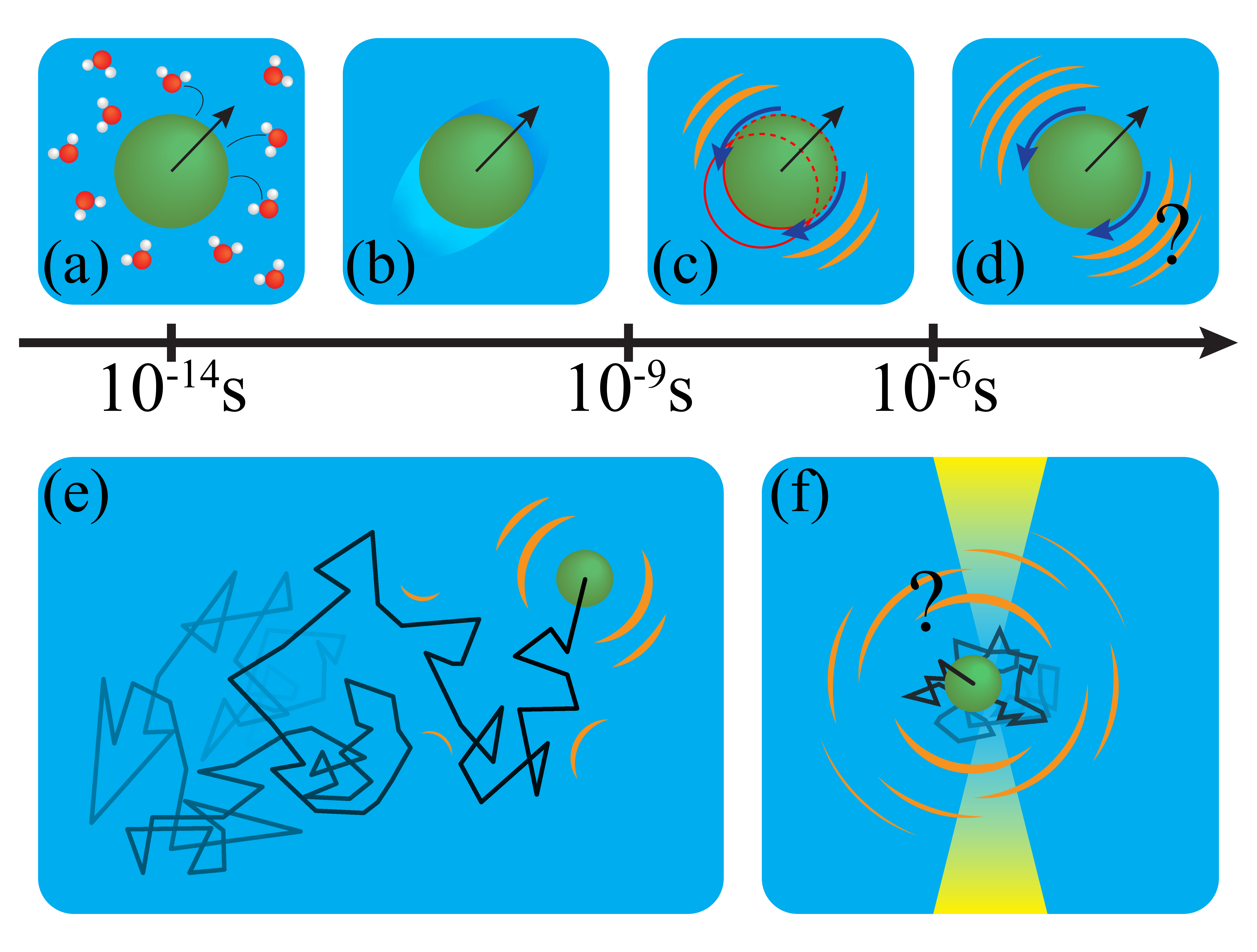}
    \caption{Illustration of colloidal dynamics across time scales. (a) The colloid first encounters the surrounding liquid via molecular collisions, beyond which the hydrodynamic limit may be reached. (b) The fluid is effectively compressible at nanoseconds (density variations denoted in color). (c) In an incompressible low-Reynolds-number flow, a colloid experiences Stokes friction (dark blue arrows), added mass due to displaced-fluid inertia (region enclosed in red), and Basset memory due to scale-free fluid-momentum diffusion (orange arcs). (d) In this work, we provide a theory incorporating Basset memory in the overdamped (inertialess) limit, which is applicable after microseconds. (e) Indeed, Basset memory contributes slowly-decaying corrections to the mean-squared displacement of a free particle. (f) In this work, we predict that Basset memory should always ultimately dominate the relaxation of trapped particles.}
    \label{fig:illust}
\end{figure}

Once a micron-sized colloid is submerged in a liquid (\eg water), its dynamics spans a rich hierarchy of physical regimes (Fig.~\ref{fig:illust}). Within tens of femtoseconds~\cite{FeynmanBOOK1964}, individual molecular collisions self-average to give rise to a fluctuating-hydrodynamic limit~\cite{BocquetCSR2010}. Within nanoseconds, the colloid becomes subject to the constraint of fluid incompressibility~\cite{ZwanzingJFM1975}. At microseconds, the inertia of the colloid becomes negligible as well~\cite{WeitzPRL1989}. However, one effect\,---\,Basset friction\,---\,contributes to power-law hydrodynamic memory of the colloid's motion, as opposed to previous effects which entail exponential decorrelation (with the above typical times). This heavy-tailed relaxation manifests measurably in the velocity autocorrelation function~\cite{PaulJPA1981} and mean-squared displacement~\cite{LukicPRL2005} of colloids. At the same time, colloidal dynamics are largely modeled using the overdamped Langevin equation, an inertia- and memoryless stochastic equation of motion. What is the fate of power-law hydrodynamic memory in the overdamped limit?

In this Letter, we show how to consistently incorporate hydrodynamic memory into the particle's overdamped Langevin equation. We identify the inverse Schmidt number\,---\,the ratio of colloid and fluid-momentum diffusivities (Eq.~\eqref{eq:epsilon})\,---\,as the small parameter that both orchestrates the exponential loss of colloid inertia as well as sets the magnitude of hydrodynamic memory. Our central result, we term the overdamped Basset-Langevin equation (Eq.~\eqref{eq:OBLE}, Fig.~\ref{fig:illust}(d)), contains the leading-order correction in Basset memory to the conventional overdamped Langevin equation for a Brownian particle. We stress that by ``inertialess'' and ``overdamped'' we mean the elimination of particle inertia; hydrodynamic memory arises from host-fluid inertia, which we do not neglect. Its fundamental consequence is that while often optically trapped particles are modeled as though they exponentially relax to steady state via Stokes friction, the Basset memory contributes to a power-law decay which ultimately overpowers the na\"ive exponential asymptotics (Eq.~\eqref{eq:RR_trap}, Fig.~\ref{fig:illust}(f)). We posit that the asymptotic decay of any observable in a colloid system, subject to arbitrary confinement and trapping force, must therefore be a power law, which our framework (presented here and in our companion article~\cite{LONG}) allows to predict analytically. Using dimensional analysis, we identify that memory is more pronounced at smaller scales and under a stronger force, offering promising avenues for controlling nanoscale systems.

\textit{Background.}---Consider a spherical colloid whose radius and mass density are, respectively, $a$ and $\rho_\p$, immersed in an unbounded viscous fluid whose thermal energy, viscosity, and mass density are, respectively, $\kt$, $\eta$, and $\rho_\f$. Suppose the particle is subjected to an external force, $\vecf$. At time $t$, denote the particle (stochastic) velocity $\vecv_t$ and position $\vecr_t$. The conventional underdamped (``inertia-full'') Langevin equation for the particle reads~\cite{ClercxPRA1992,SchussBOOK2010} 
\begin{equation}
    \frac43\pi a^3\rho_\p\frac{\d\vecv_t}{\d t}=-6\pi \eta a \vecv_t+\vecf(\vecr_t,t)+\sqrt{2\kt(6\pi \eta a)}\bxi_t,\label{eq:underdamped}
\end{equation}
where the thermal force satisfies $\langle \bxi_t\bxi_{t'}\rangle=\delta(t-t')\vecI$, with $\vecI$ the $3\times3$ unit matrix. By eliminating inertia on the left-hand side, Eq.~\eqref{eq:underdamped} can be rigorously brought to the overdamped (colloid-inertialess) limit~\cite{SchussBOOK2010,GardinerBOOK2009}, for which
\begin{equation}
    \frac{\d\vecr_t}{\d t}=\frac D\kt\vecf(\vecr_t,t)+\sqrt{2D}\bxi_t,\label{eq:overdamped}
\end{equation}
where $D=\kt/(6\pi\eta a)$ is the particle's diffusion constant.

Equation~\eqref{eq:underdamped} is incomplete; it implies that, for example, the velocity autocorrelation of a free particle ($\vecf=\mathbf0$) would decay exponentially, $\langle\vecv_t\vecv_0\rangle\sim e^{-(9\eta/2a^2\rho_\p)t}\,\vecI$. However, it is numerically~\cite{AdlerPRL1967,AdlerPRA1970}, theoretically~\cite{ErnstPRL1970,HinchJFM1975}, and experimentally~\cite{BoonPLA1976,BouillerJPF1978,PaulJPA1981} established that hydrodynamic memory contributes to a power-law decay of the velocity autocorrelation function, $\langle\vecv_t\vecv_0\rangle\sim t^{-3/2}\vecI$. 

The complete form of Eq.~\eqref{eq:underdamped} for the motion of the colloid is constructed as follows. For low-Reynolds-number unsteady particle motion in the incompressible flow limit, the viscous friction comprises the familiar steady Stokes drag, an added-mass contribution associated with the inertia of the displaced fluid, and, crucially for our work, a power-law hydrodynamic (Basset) memory term arising from the self-similar diffusion of momentum through the fluid~\cite{LandauBOOK1987}. Upon including thermal forces from the fluid, the complete equation of motion of the particle is the fluctuating Basset-Boussinesq-Oseen equation~\cite{ClercxPRA1992},
\begin{align}
    \frac43\pi a^3\rho_\p\frac{\d\vecv_t}{\d t}=&-6\pi \eta a \vecv_t-\frac23\pi a^3\rho_\f\frac{\d\vecv_t}{\d t}\nonumber\\& -6a^2\sqrt{\pi\rho_\f\eta}\int_{-\infty}^t\frac{\d \vecv_{t'}}{\sqrt{t-t'}}\nonumber\\&+\vecf(\vecr_t,t)+\sqrt{2\kt(6\pi \eta a)}\bta_t,\label{eq:FBBOE}
\end{align}
where the fluid is otherwise in thermal equilibrium so it satisfies the fluctuation-dissipation theorem~\cite{HaugeJSP1973}, requiring $\langle \bta_t\bta_{t'}\rangle=\{\delta(t-t')-(1/4)[\rho_\f a^2/(\pi\eta)]^{1/2}|t-t'|^{-3/2}\}\vecI$.
The integral term in Eq.~\eqref{eq:FBBOE}, the Basset memory, is responsible for the power-law decay of velocity autocorrelations~\cite{ErnstPRL1970,HinchJFM1975}. We now ask: how to consistently bring Eq.~\eqref{eq:FBBOE} to an overdamped (inertialess) form, in the presence of hydrodynamic memory? 

\textit{Nondimensionalization.}---We proceed to nondimensionalize Eq.~\eqref{eq:FBBOE}, which will aid in identifying the parameter controlling the magnitude of memory. The conventional overdamped equation, Eq.~\eqref{eq:overdamped}, models a particle exploring a force landscape $\vecf$ via Brownian motion, with a diffusion constant $D$. The particle size, $a$, does not appear in the equation outside of $D$; the relevant length scale is instead that over which the force varies, which we denote $w$. (In the absence of a force\,---\,a scale-free problem\,---\,one may equivalently choose $w=a$.) For a time-independent conservative force, by the equipartition theorem~\cite{PathriaBOOK2011}, $\lim_{t\to\infty}\langle\vecr_t\vecf(\vecr_t)\rangle=-\kt\vecI$, which implies that the force magnitude scales as $\kt/w$. Under such a force magnitude, the particle velocity scales as $(6\pi\eta a)^{-1}\times\kt/w=D/w$. We suppose that if the external driving force changes in time, it will change during the time scale of colloidal diffusion, $w^2/D$. Since the noise was defined to have dimensions of inverse square-root time, the colloidal-diffusion time scale sets its magnitude as $D^{1/2}/w$. We thus nondimensionlize as follows:\footnote{Mind that, throughout, $T$ is dimensionless time, and $\kt$ is temperature.}
\begin{gather}
    T:=\frac{Dt}{w^2},\quad\vecR:=\frac\vecr w,\quad\vecV:=\frac{w\vecv}D,\nonumber\\\quad\vecF(\vecR,T):=\frac{w\vecf(w\vecR,w^2T/D)}\kt,\quad(\bTa,\bXi):=\frac{w(\bta,\bxi)}{D^{1/2}}.\label{eq:nondim}
\end{gather}

In terms of these dimensionless variables, Eq.~\eqref{eq:FBBOE} becomes
\begin{equation}
    \kappa\epsilon\frac{\d\vecV_T}{\d T}=-\vecV_T-\epsilon^{1/2}\int_{-\infty}^T\frac{\d \vecV_T'}{\sqrt{T-T'}}+\vecF(\vecR_T,T)+\sqrt{2}\bTa_T,\label{eq:FBBOE_nondim}
\end{equation}
with the nondimensional (colored) noise $\langle \bTa_T\bTa_{T'}\rangle=[\delta(T-T')-(\epsilon^{1/2}/4)|T-T'|^{-3/2}]\,\vecI$.
In Eq.~\eqref{eq:FBBOE_nondim} we identified two parameters. One is set by the ratio of colloid and fluid densities, $\kappa:=(2\pi/9)(\rho_\p/\rho_\f+1/2)$. Often experimentalists utilize density-matched colloids ($\rho_\p=\rho_\f$) to eliminate gravitational effects, in which case $\kappa=\pi/3$. Thus, we regard $\kappa$ as an order-$1$ parameter. The other, more important, parameter is
\begin{equation}
    \epsilon:=\frac1\pi\frac{\rho_\f D}{\eta}\frac{a^2}{w^2}.\label{eq:epsilon}
\end{equation}
It is related to the (inverse) Schmidt number, $\mathrm{Sc}^{-1}=\rho_\f D/\eta$, comparing the diffusivity of the colloid with that of the diffusivity of momentum within the fluid. For a micron-sized colloid in water at room temperature, $\mathrm{Sc}^{-1}\sim10^{-6}$. 

The nondimensional Eq.~\eqref{eq:FBBOE_nondim} makes the hierarchy of terms explicit. A small $\epsilon$ is ultimately equivalent to a long-time expansion, by which we mean that we are concerned with the time it takes the particle to explore the force landscape, $t\sim w^2/D$, as opposed to the time it takes the fluid momentum to diffuse the scale of the particle, $a^2/(\eta/\rho_\f)$; no other time scales are present in Eq.~\eqref{eq:FBBOE}. Clearly, the dominant contribution to the particle drag is the steady-Stokes friction, the second term in Eq.~\eqref{eq:FBBOE_nondim}. The next-order frictional term is the Basset memory, entailing a fractional power of the time-scale ratio that originates from diffusion of fluid vorticity. Lastly, note that the left-hand side of Eq.~\eqref{eq:FBBOE_nondim} contains both particle and displaced-fluid inertias via $\kappa$, proportional to an even-higher order in the time-scale ratio. With this setup in mind, we provide the main result and key conceptual takeaways.

\textit{Main result.}---The above hierarchy highlights two underappreciated concepts. First, we see two limits in Eq.~\eqref{eq:underdamped}: (i) The memory term, in fact, dominates over inertia at late times, so one must not ignore it. (ii) Even if one wishes to account for the subdominant inertia of the particle, it is necessary to include the added-mass of the surrounding incompressible flow, which is of the same order of magnitude. Thus, while the underdamped Langevin model may be convenient and minimal, it is rather unrealistic for modeling colloids. Second, importantly, this hierarchy implies that an overdamped (inertialess) limit may contain hydrodynamic memory still: at late times ($\epsilon\ll1$) inertia becomes negligible before memory does. As we show next, although memory is identified with a small (order-$\epsilon^{1/2}$) magnitude, it nonetheless dominates at late times.

With this acquired understanding, the next step is a small-$\epsilon$ expansion of Eq.~\eqref{eq:FBBOE_nondim} to eliminate the velocity variable and obtain the overdamped (inertialess) limit. In our companion article~\cite{LONG}, we present a detailed and rigorous derivation based on stochastic Taylor expansions~\cite{SchussBOOK2010,KloedenBOOK1992,GardinerBOOK2009}. For completeness, in the \hyperref[EndMatter]{End Matter}, we present a derivation based on a Fourier transform. The result is a closed-form equation for $\d\vecR_T/\d T$, correct to order $\epsilon^{1/2}$, which we term the overdamped Basset-Langevin equation,
\begin{align}
	\frac{\d\vecR_T}{\d T}=&\vecF(\vecR_T, T)-\epsilon^{1/2}\int_{-\infty}^T\frac{\d \vecF(\vecR_{T'},T')}{\sqrt{T-T'}}\nonumber\\&+\sqrt2\,\bXi_T+\sqrt{2}\,\frac{\epsilon^{1/2}}8\int_{-\infty}^\infty\frac{\bXi_{T'}\d T'}{|T-T'|^{3/2}},\label{eq:OBLE}
\end{align}
where $\langle\bXi_T\bXi_{T'}\rangle=\delta(T-T')\,\vecI$ is white noise. Equation~\eqref{eq:OBLE} is the central result of this Letter. (The dimensional form of Eq.~\eqref{eq:OBLE} and of all later equations, as well as a clarification on the stochastic integral $\int\d \vecF(\vecR_{T'})$, are provided in the End Matter.) In addition to the (steady-Stokes) first and (white-noise) third terms on the right-hand side of Eq.~\eqref{eq:OBLE}, which are present in the conventional overdamped Langevin equation, Eq.~\eqref{eq:overdamped}, we obtain two new integral terms, encoding the Basset memory of the past applied forces and noises, respectively.

The identification of a small parameter $\epsilon$ permits convenient expansion schemes to find the positional evolution of a colloid. Furthermore, much of soft- and active-matter modeling involves the overdamped Langevin equation~\cite{VitelliBOOK2024} as the inertia of micron-sized particles is negligible. Equation~\eqref{eq:OBLE} is thus a closed form expression for the overdamped dynamics of interest, which accounts for Basset memory. In the remainder of the Letter, we assess how dominant is the Basset memory through two examples.

\textit{Example: Free particle.}---Consider first the much studied free-particle case, $\vecF=\mathbf0$. Via a Fourier transform, one may confirm that the full Eq.~\eqref{eq:FBBOE_nondim} provides the well-known power-law decay in velocity autocorrelation, $\langle\vecV_T\vecV_0\rangle=(\epsilon^{1/2}/2)T^{-3/2}\vecI$, up to order $\epsilon$. More pertinent to our work is the impact of hydrodynamic correlations on particle diffusion. From Eq.~\eqref{eq:OBLE}, one readily obtains the mean-squared displacement,
\begin{equation}
    \langle(\vecR_T-\vecR_0)(\vecR_T-\vecR_0)\rangle=2T\left(1-\frac{2\epsilon^{1/2}}{T^{1/2}}\right)\vecI,\label{eq:MSD_free}
\end{equation}
up to order-$\epsilon$ corrections. This result coincides with past theory~\cite{WeitzPRL1989} and experiment~\cite{LukicPRL2005,HuangNP2011}, showing that hydrodynamic memory impacts the diffusion of colloids in a measurable way. Since the problem of particle diffusion is scale free, like the diffusion of fluid momentum, the hydrodynamic memory as a result of the latter does not dominate the asymptotics. (See illustration in Fig.~\ref{fig:illust}(e), wherein the colloids traverses further than the vorticity's backaction.) Nevertheless, due to the  small power-law, the Basset correction is persistent, only decaying to $1\%$ past $T\gtrsim10^4$~\cite{WeitzPRL1989}. In the following example, hydrodynamic memory will, in fact, dominate relaxation.

\begin{figure}
    \centering
    \includegraphics[width=0.99\linewidth]{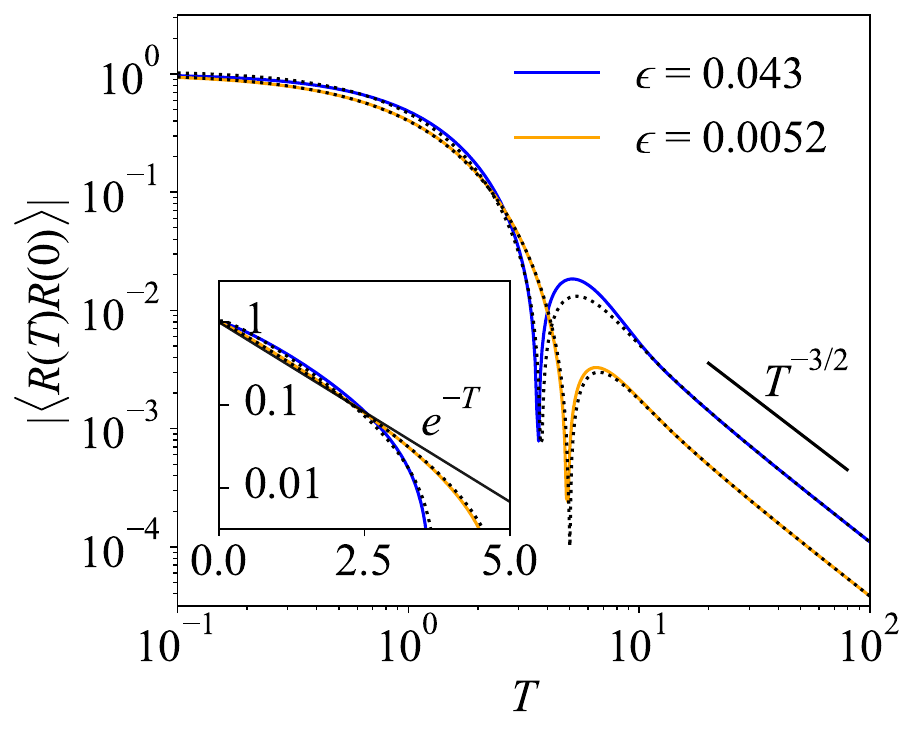}
    \caption{Nondimensionalized positional autocorrelation function of a harmonically trapped colloid versus time in a log-log plot. The solid curves are numerical solutions to the full fluctuating Basset-Boussinesq-Ossen equation, Eq.~\eqref{eq:FBBOE_nondim}, evaluated through a Fourier transform (Eq.~\eqref{eq:noisy_BBO_nondim_FT}). Color corresponds to the indicated $\epsilon$ values (Eq.~\eqref{eq:epsilon}), which we chose to correspond to the two experiments in Ref.~\cite{FranoschNATURE2011}. The directly adjacent black dashed curves are our late-time prediction, Eq.~\eqref{eq:RR_trap}, found from the overdamped Basset-Langevin equation (Eq.~\eqref{eq:OBLE}). Inset: A log-linear plot of the early-time, primarily exponential decay of the positional autocorrelation function. We clearly see the crossover from primarily exponential to powerlaw-$3/2$-dominated decay.}
    \label{fig:trap-linResp}
\end{figure}

\textit{Example: Trapped particle.}---Consider now a trapped particle, $\vecF(\vecR)=-\vecR$. To our knowledge, very limited theoretical literature is available for treating this problem~\cite{ClercxPRA1992,LukicPRE2007}, primarily aimed at estimating the mean-squared displacement and the velocity autocorrelation function of such trapped particles. Here we provide a probe for the importance of Basset memory\,---\,the positional autocorrelation function. Once again, we readily obtain it from Eq.~\eqref{eq:OBLE},
\begin{equation}
    \langle\vecR_T\vecR_0\rangle=\left(e^{-T}-\frac{\epsilon^{1/2}}{2T^{3/2}}\right)\vecI,\label{eq:RR_trap}
\end{equation}
up to order-$\epsilon$ and $\epsilon^{1/2}/T^{5/2}$ corrections. We plot Eq.~\eqref{eq:RR_trap} along with a numerical evaluation of $\langle\vecR_T\vecR_0\rangle$ from the complete Eq.~\eqref{eq:FBBOE_nondim} in Fig.~\ref{fig:trap-linResp}.

Equation~\eqref{eq:RR_trap} encodes a crossover: The leading-order term is an exponential decay to the trap center as a consequence of the steady-Stokes friction, $6\pi\eta a\,\d\vecr_t/\d t=-(\kt/w^2)\vecr_t$. The second term in \eqref{eq:RR_trap} arises from the Basset memory of previously applied forces. Notably, the memory gives rise to an anticorrelation, which can be understood by the increase in fluctuations due to the additional noise memory term in Eq.~\eqref{eq:OBLE}, which is absent in Eq.~\eqref{eq:overdamped}.

The key finding is that the order-$\epsilon^{1/2}$ Basset relaxation will eventually dominate over the order-$1$ term, as it decays as a power law, slower than an exponential. Even for a very small $\epsilon$, the crossover time $T\sim\ln(\epsilon^{-1/2})$ from Stokes (exponential) to Basset (power-law) relaxation is merely of order $\simeq10$ for a micron-sized particle ($\epsilon\sim10^{-6}$), making it realistic for measurement.\footnote{We note that in the case of a free particle, where, qualitatively, $\langle\vecV_T\vecV_0\rangle=[(\kappa\epsilon)^{-1}e^{-T/(\kappa\epsilon)}+(\epsilon^{1/2}/2)T^{-3/2}]\vecI$, one must employ intricate tools to observe the inertial regime~\cite{HuangNP2011}, as the crossover time is very small, $T\sim\epsilon\ln(\epsilon^{-1/2})\ll1$.} Indeed, in Ref.~\cite{FranoschNATURE2011},
for a micron-sized particle in a very stiff trap, the positional autocorrelation of a trapped colloids has been measured experimentally to high resolution; see End Matter. The authors of Ref.~\cite{FranoschNATURE2011} observe such a crossover that culminates in a $t^{-3/2}$ positional anticorrelation, which was argued for heuristically, based on the color of the noise spectrum in the full Eq.~\eqref{eq:FBBOE}. Equation~\eqref{eq:OBLE}, on the other hand, provides rigorous grounds to understand this phenomenon. Indeed, in accord with Eq.~\eqref{eq:RR_trap}, the correlation-to-anticorrelation crossover in Ref.~\cite{FranoschNATURE2011} occurred at about $10$ times the Stokesian relaxation time. (In fact, the exponent does not seem to fit well the exact autcorrelation function, since higher-order (order-$\epsilon^{1/2}/T^{5/2}$ and others) hydrodynamic corrections may already cause a deviation at $T\sim1$; \cf Fig.~\ref{fig:trap-linResp}.) Furthermore, the $t^{-3/2}$ power law was observed at an earlier time and thus was better resolved for a particle immersed in a less viscous fluid (acetone versus water).
This latter result is in agreement with Eq.~\eqref{eq:epsilon}, supporting that Basset corrections become more pronounced in lower-viscosity host fluids. 

\textit{Discussion.}---In this work, we investigated the fate of hydrodynamic memory in the overdamped (inertialess) limit of colloidal dynamics. We isolated the small nondimensional parameter $\epsilon$, Eq.~\eqref{eq:epsilon}, orchestrating the crossover from the underdamped to overdamped limits. It is given by the ratio between the time scales for the fluid's momentum diffusion and of the particle's diffusive exploration\,---\,the former being substantially shorter. With $\epsilon$, starting from the full fluctuating Basset-Boussinesq-Oseen equation, Eq.~\eqref{eq:FBBOE_nondim}, we obtain the overdamped Basset-Langevin equation, Eq.~\eqref{eq:OBLE}\,---\,our central result, which contains the conventional force-induced drift and noise terms, as well as (Basset) memory terms of the force and noise histories. Since the result is given in closed form, it permits the treatment of overdamped colloidal dynamics while taking into account the hydrodynamic memory. 

The examples provided above support the fact that Basset memory is observable and asymptotically dominates relaxation even in the inertialess limit. As a particle explores the force landscape $\vecF$ during $T\sim1$ ($t\sim w^2/D$), one may expect that by $T\sim10$ the particle mostly explores the basin of the external force, where $\vecF\sim-\vecR$. Thus, we posit that the crossover from Stokesian exponential decay to Basset power-law decay occurs universally across all relaxing colloids.

We made two assumptions in our starting point, Eq.~\eqref{eq:FBBOE}. First, the flow is at low Reynolds number. Since the particle moves (and thus the fluid is carried along) with typical velocity $(6\pi\eta a)^{-1}\times\kt/w=D/w$ a distance of $w$, the corresponding Reynolds number is $\mathrm{Re}=\rho_\f (D/w) w/\eta=\mathrm{Sc}^{-1}\sim\epsilon\ll1$. 
In our companion article, using the reciprocal theorem~\cite{TurkJFM2025}, we show that the nonlinear advection term in the Navier-Stokes equation would have contributed to an order-$\epsilon^{3/2}$ term, which is thus even less important than inertia at late times. Second, the flow is incompressible. As the particle is in perpetual Brownian motion for $\kt>0$, at very short times, the fluid will, in fact, be compressible. However, the corrections from fluid compressibility decay exponentially after time $\sim a/c$, where $c$ is the speed of sound in the fluid~\cite{ZwanzingJFM1975}. For a micron-sized colloid in water at room temperature, $(a/c)/(\rho_\f a^2/\eta)\sim10^{-3}$, meaning that the incompressible approximation applies long before the momentum has diffused through the fluid, let alone the scale of the colloid. Thus, both assumptions are reasonable at late times. Lastly, nowhere in our work have we considered the case of $\kappa:=(2\pi/9)(\rho_\p/\rho_\f+1/2)\gg1$ or $\kappa\ll1$, as the particle would sink or float. The more general framework we present in our companion article can be used to describe the diffusion of a particle in those scenarios~\cite{LONG}. Ultimately, tuning $\kappa$ would only affect the time at which inertia can be eliminated, which is negligible in the vast majority of colloidal experiments.

The effect of hydrodynamic memory surviving in the overdamped-particle limit that we describe here appears to be general. Yet in only few experiments has it been possible to resolve the memory in the overdamped positional relaxation~\cite{FranoschNATURE2011}. This difficulty arises since the power-law decay dominates over the exponential decay asymptotically, so it may not occur sufficiently early to be measurable amid noise. Indeed, it took half a decade to report a statistically significant power-law decay of velocity autocorrelations~\cite{PaulJPA1981} since the first experimental attempt~\cite{BoonPLA1976}. Likewise, for overdamped observables, we are only aware of Ref.~\cite{FranoschNATURE2011} where high-resolution measurements of a colloid's positional correlations were made. Although the crossover occurs at $T\lesssim10$, the autocorrelations have by then decayed three order of magnitude, requiring high data quality to statistically significantly resolve the power-law decay; such precision may exceed that available in most colloidal-trapping experiments. 

While the host-fluid density, $\rho_\f$, viscosity $\eta$, and temperature $\kt$ can only be tuned quantitatively (within one order of magnitude), we note that $\epsilon\sim (\rho_\f\kt/\eta^2)\times 1/a\times(a/w)^2$ can be substantially increased by considering smaller particles (decreasing $a$) and stronger forcings (decreasing $w/a$), where the others parameters are material properties and less amendable for substantial modification. So far, we considered $a,w\sim10^{-6}\,\mathrm{m}$. However, if one instead considers a quantum dot ($a\sim10^{-9}\,\mathrm{m}$) under a strong confining potential (say, $a/w\sim10$), then $\epsilon^{1/2}\sim10^{-1}$, increasing the importance of the Basset memory in the relaxational dynamics of the particles. Given the significant improvement of tracking and trapping techniques for nanometer-sized particles, such as quantum dots~\cite{McHaleNL2007,RifeLANGMUIR2009,DuCC2013} and biomolecules~\cite{ShuBIORXIV26,XiangNM2020,CohenPNAS2006}, we are optimistic that hydrodynamic memory can be resolved experimentally more extensively in the near future. As experimentalists seek to better control nanoscale structures, we posit that this long-tailed memory may contribute to, \eg enhance transport~\cite{SeylerPRR2019} and induce synchronization~\cite{KurebayashiEPL2012}.

Lastly, the effect of confinement would also be interesting to explore. Relaxation processes in the presence of walls are faster but still power laws, $t^{-5/2}$~\cite{FelderhofJPCB2005,JeneyPRL2008,FranoschPRE2009}. It may be instructive to investigate the longevity of hydrodynamic memory in the presence of momentum scatterers and absorbers, and in the presence of swimmers, or other active matter, contributing to excess fluctuations in the surrounding fluid. Theoretical investigation of effects like these would assist in assessing the importance of hydrodynamic memory inside cells and bacterial suspensions.

\acknowledgements

B.S. was supported by the Princeton Center for Theoretical Science and in part by the Center for the Physics of Biological Function at Princeton University. H.A.S and G.T. acknowledge support from the U.S. National Science Foundation via grant
No. CBET-2246791 and through the Princeton Center for Complex Materials (DMR-2011750).

\begin{widetext}
\subsection{End Matter}\label{EndMatter}

\textit{Derivation of main result.}---Here, owing to Eq.~\eqref{eq:FBBOE} being linear in $\vecv_t$ and $\bta_t$ (aside from the force term), we may derive our main result Eq.~\eqref{eq:OBLE} from Eq.~\eqref{eq:FBBOE_nondim} via a Fourier transform. We will work in the conventions $\tilde X(\Omega)=\int_{-\infty}^\infty X_Te^{-\i\Omega T}\d T$ and $X_T=(2\pi)^{-1}\int_{-\infty}^\infty \tilde X(\Omega)e^{\i\Omega T}\d \Omega$ for a stochastic time-dependent quantity $X_T$. Upon Fourier transforming Eq.~\eqref{eq:FBBOE_nondim},\footnote{We used the following properties: $\int_{-\infty}^\infty e^{-\i\Omega T}\d \vecV_T=\i\Omega\tilde \vecV(\Omega)$, $\int_0^\infty e^{-\i\Omega T}T^{-1/2}\d T=(\pi/2)^{1/2}|\Omega|^{-1/2}(1-\i\sgn\Omega )$, and $\int_{-\infty}^{\infty}e^{-\i\Omega T}|T|^{-3/2}\d T=-2(2\pi|\Omega|)^{1/2}$, along with $(\d \vecV_T/\d T)\d T=\d\vecV_T$.} we find
\begin{equation}
    \i\kappa\epsilon\Omega\tilde \vecV(\Omega)=-\tilde\vecV(\Omega)-\sqrt{\frac{\pi\epsilon|\Omega|}2}(1+\i\sgn\Omega )\tilde\vecV (\Omega)+\int_{-\infty}^\infty\vecF(\vecR_T,T)e^{-\i\Omega T}\d T+\sqrt{2}\tilde\bTa(\Omega),\label{eq:noisy_BBO_nondim_FT}
\end{equation}
with the Fourier-transformed nondimensional (colored) noise\footnote{Of all expressions to far, it is easiest to confirm the fluctuation-dissipation theorem from this equation: Indeed the fluctuation magnitude is given by~\cite{PathriaBOOK2011} $\langle \tilde\bTa(\Omega)\tilde\bTa(\Omega')\rangle=2\pi\delta(\Omega+\Omega')\vecI\Re[\tilde\Gamma(\Omega)]$, where $\Re[\tilde\Gamma(\Omega)]$ is the real part of the friction matrix, \ie the prefactor in front of $\tilde\vecV(\Omega)$ in Eq.~\eqref{eq:noisy_BBO_nondim_FT}, $\Gamma(\Omega)=1+(\pi\epsilon|\Omega|/2)^{1/2}(1+\i\sgn\Omega)-\i\kappa\epsilon\Omega$.}  $\langle \tilde\bTa(\Omega)\tilde\bTa(\Omega')\rangle=2\pi\delta(\Omega+\Omega')[1+(\pi\epsilon|\Omega|/2)^{1/2}]\vecI$. This is an algebraic equation in $\tilde\vecV(\Omega)$, which we may thus isolate to order $\epsilon^{1/2}$,
\begin{equation}
    \tilde \vecV(\Omega)=\left[1-\sqrt{\frac{\pi\epsilon|\Omega|}2}(1+\i\sgn\Omega)\right]\int_{-\infty}^\infty\vecF(\vecR_T,T)e^{-\i\Omega T}\d T+\sqrt2\tilde\bTa'(\Omega),\label{eq:V_nondim_FT_interrim}
\end{equation}
where we defined the noise $\tilde\bTa'(\Omega)=[1-(\pi\epsilon|\Omega|/2)^{1/2}(1+\i\sgn\Omega)]\tilde\bTa(\Omega)$. To order $\epsilon^{1/2}$, $\tilde\bTa'(\Omega)$ satisfies $\langle \tilde\bTa'(\Omega)\tilde\bTa'(\Omega')\rangle=2\pi\delta(\Omega+\Omega')[1-(\pi\epsilon|\Omega|/2)^{1/2}]\vecI$. Recalling that the Fourier transform of white noise is $\langle \tilde\bXi(\Omega)\tilde\bXi(\Omega')\rangle=2\pi\delta(\Omega+\Omega')\vecI$,
we replace $\tilde\bTa'(\Omega)$ in Eq.~\eqref{eq:V_nondim_FT_interrim} with white noise as
\begin{equation}
    \tilde \vecV(\Omega)=\left[1-\sqrt{\frac{\pi\epsilon|\Omega|}2}(1+\i\sgn\Omega)\right]\int_{-\infty}^\infty\vecF(\vecR_T,T)e^{-\i\Omega T}\d T+\sqrt2\left(1-\frac12\sqrt{\frac{\pi\epsilon|\Omega|}2}\right)\tilde\bXi(\Omega).\label{eq:V_nondim_FT}
\end{equation}
An inverse Fourier transform of Eq.~\eqref{eq:V_nondim_FT} yields Eq.~\eqref{eq:OBLE}. Importantly, Eq.~\eqref{eq:V_nondim_FT} satisfies the fluctuation-dissipation theorem.

\textit{Stochastic force increment.}---Note that Eq.~\eqref{eq:OBLE} involves the quantity $\d \vecF(\vecR_T,T)$. Since $\d\vecR_T$ involves a Wiener increment, we must evaluate $\d \vecF(\vecR_T,T)$ carefully using the It\^o lemma~\cite{SchussBOOK2010}. Since $\d \vecF(\vecR_T,T)$ appears within an order-$\epsilon^{1/2}$ term, we may compute it only using the leading-order in Eq.~\eqref{eq:OBLE},
\begin{equation}
    \frac{\d \vecF(\vecR_{T},T)}{\d T}=\frac{\partial \vecF(\vecR_{T},T)}{\partial T}+[\vecF(\vecR_T,T)\cdot\grad ]\vecF(\vecR_T,T)+\sqrt2(\bXi_T\cdot\grad )\vecF(\vecR_T,T)+D\nabla^2\vecF(\vecR_T,T),
\end{equation}
where the last term arose from the It\^o lemma. Therefore, the fully explicit version of our central result, Eq.~\eqref{eq:OBLE}, reads
\begin{align}
	\frac{\d\vecR_T}{\d T}=&\,\vecF(\vecR_T,T)-\epsilon^{1/2}\int_{-\infty}^T\frac{[\vecF(\vecR_{T'},T')\cdot\grad ]\vecF(\vecR_{T'},T')+\nabla^2\vecF(\vecR_{T'},T')+\partial\vecF(\vecR_{T'},T')/\partial T'}{\sqrt{T-T'}}\d T'\nonumber\\&+\sqrt2\bXi_T+\sqrt{2}\epsilon^{1/2}\left[\frac18\int_{-\infty}^\infty\frac{\bXi_{T'}\d T'}{|T-T'|^{3/2}}-\int_{-\infty}^T\frac{(\bXi_{T'}\cdot\grad)\vecF(\vecR_{T'},T')}{\sqrt{T-T'}}\d T'\right].\label{eq:OBLE_explicit}
\end{align}

\textit{Equations with dimensions.}---Here, we bring back dimensions into the nondimensional equations we have written in the main text. Equation~\eqref{eq:OBLE}:
\begin{equation}
	\frac{\d\vecr_t}{\d t}=\frac D\kt\vecf(\vecr_t,t)-\frac D\kt\sqrt{\frac{\rho_\f a^2}{\pi\eta}}\int_{-\infty}^t\frac{\d \vecf(\vecr_{t'},t')}{\sqrt{t-t'}}+\sqrt{2D}\bxi_T+\sqrt{2D}\sqrt{\frac{\rho_\f a^2}{\pi\eta}}\frac18\int_{-\infty}^\infty\frac{\bxi_{T'}\d t'}{|t-t'|^{3/2}}.\label{eq:overdamped_BBO}
\end{equation}
Equation~\eqref{eq:MSD_free} (in agreement with Ref.~\cite{WeitzPRL1989}):
\begin{equation}
    \langle(\vecr_t-\vecr_0)(\vecr_t-\vecr_0)\rangle=2Dt\left(1-2\sqrt{\frac{\rho_\f a^2}{\pi\eta}}\frac1{t^{1/2}}\right)\vecI.
\end{equation}
Equation~\eqref{eq:RR_trap}:
\begin{equation}
    \langle\vecr_t\vecr_0\rangle=w^2\left(e^{-Dt/w^2}-\frac{w^2}{2D}\sqrt{\frac{\rho_\f a^2}{\pi\eta}}\frac1{t^{3/2}}\right)\vecI.
\end{equation}
Equation~\eqref{eq:OBLE_explicit}:
\begin{eqnarray}
	\frac{\d\vecr_t}{\d t}&=&\frac D\kt\vecf(\vecr_t)-\frac {D}\kt \sqrt{\frac{\rho_\f a^2}{\pi\eta}}\int_{-\infty}^t\frac{D[(\kt)^{-1}\vecf(\vecr_{t'},t')\cdot \grad_\vecr]\vecf(\vecr_{t'},t')+D\nabla^2_\vecr\vecf(\vecr_{t'},t')+\partial\vecf(\vecr_{t'},t')/\partial t'}{\sqrt{t-t'}}\d t'
	\nonumber\\&&+\sqrt{2D}\bxi_t+\sqrt{2D}\sqrt{\frac{\rho_\f a^2}{\pi\eta}}\left[\frac18\int_{-\infty}^\infty\frac{\bxi_{t'}\d t'}{|t-t'|^{3/2}}-\frac {D}\kt\int_{-\infty}^t\frac{(\bxi_{t'}\cdot\grad_\vecr)\vecf(\vecr_{t'},t')}{\sqrt{t-t'}}\d t'\right].
\end{eqnarray}

\textit{Experimental parameters.}---Here, we summarize the parameters in the experiment of Ref.~\cite{FranoschNATURE2011}. The trap stiffness is $k=2.05\cdot10^{-4}\,\mathrm{N}/\mathrm{m}$, meaning that at room temperature, the force landscape varies across $w=(\kt/k)^{1/2}=4.49\cdot10^{-9}\,\mathrm{m}$. The colloid radius, on the other hand, is $a=1.45\cdot10^{-6}\,\mathrm{m}$. Therefore, although $\mathrm{Sc}^{-1}\ll1$, $(a/w)^2\gg1$, meaning that $\epsilon=\mathrm{Sc}^{-1}(a/w)^2$ (Eq.~\eqref{eq:epsilon}) can be increased beyond the inverse Schmidt number. The fluid-momentum diffusion time $\tau_\mathrm{f}:=\rho_\f a^2/\eta$ is $\tau_\mathrm{f}=2.3\cdot10^{-6}\,\mathrm{s}$ in water (and $\tau_\mathrm{f}=5.1\cdot10^{-6}\,\mathrm{s}$ in acetone). The relaxation time of the stiff trap, $\tau_\mathrm{p}:=6\pi\eta a/k=w^2/D$, was $\tau_k=1.4\cdot10^{-4}\,\mathrm{s}$ ($\tau_k=3.8\cdot10^{-5}\,\mathrm{s}$). Thus, according to Eq.~\eqref{eq:epsilon}, $\epsilon=\tau_\mathrm{f}/(\pi\tau_\mathrm{p})$ is $\epsilon=0.0052$ ($\epsilon=0.043$).

\end{widetext}

\end{document}